# Slightly broken icosahedral symmetry advances Thomson problem

D.S. Roshal, A.E. Myasnikova, S.B. Rochal

Faculty of Physics, Southern Federal University, 5 Zorge str., 344090, Rostov-on-Don, Russia

- We break the icosahedral symmetry of Caspar and Klug model of viral capsids.
- Simplest distortion of global icosahedral arrangement in 2D nanocrystals is found.
- Trial 2D spherical structures obtained are close to the lowest energy Thomson ones.
- A new way to combine the local hexagonal order and spherical geometry is proposed.
- List of Thomson structures with the lowest seen energies is essentially updated.

To advance Thomson problem we generalize physical principles suggested by Caspar and Klug (CK) to model icosahedral capsids. Proposed simplest distortions of the CK spherical arrangements yield new-type trial structures very close to the lowest energy ones. In the region $600 \leq N \leq 1000$, where $N$ is the number of particles in the structure, we found 40 new spherical crystals with the lowest ever seen energies and curvature-induced topological defects being not the well-known elongated scars but flatten pentagons. These crystals have $N$ values prohibited in the CK model and demonstrate a new way to combine the local hexagonal order and spherical geometry.

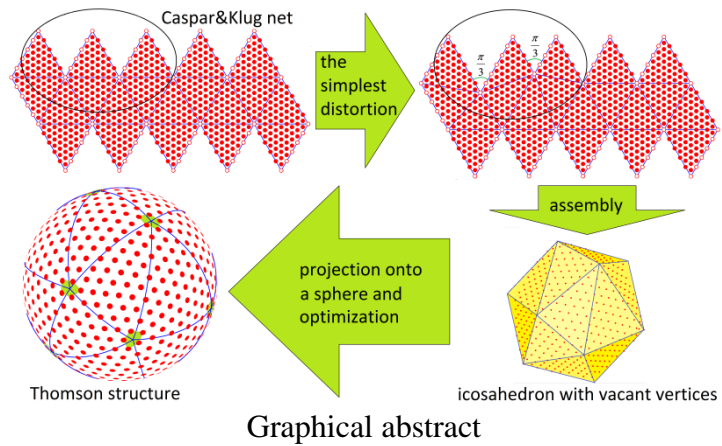

Graphical abstract

Self-organization of repelling particles retained on a spherical surface is under discussion for more than a century and is called Thomson problem [1] after J.J. Thomson suggested his model of atom 110 years ago. Now we know that Thomson problem arises on different levels of the matter self-organization. Arrangements in multi-electron bubbles in superfluid helium [2] almost perfectly correspond to structures formed by charged particles in the frame of the problem. Some of the Thomson structures (TSs) are similar to spherical viral capsids [3]. Simplest repelling potentials $1/r_{ij}^\alpha$, where $r_{ij}$ is the distance between $i^{th}$ and $j^{th}$ particles and $\alpha \geq 1$, are used to describe colloidal crystals occurring at the interface between two fluids [4-7] and porous dried colloidosomes [8-10]. If $\alpha \to \infty$ then the TSs tend to densest packings of spherical caps on a sphere. This problem is named after Tammes [11] and has a significant number of applications [12].

Experimental and theoretical studies of spherical crystals and physical properties of 2D order [4, 7, 13-15] support a new wave of interest in the Thomson problem in the past years [16-19]. The low-energy structures formed by repelling particles retained on spherical and other curved surfaces are actively used to study and classify the curvature-induced extended topological defects and reactions between them [7, 18, 20] in 2D colloidal crystals with unconventional geometry. However, the classical spherical TSs corresponding to the global energy minima are also very interesting [6, 21-23]. Their search is a challenging work since the equilibrium energies of structures corresponding to global and local-minima are very close. Moreover, the difference between the equilibrium energies is strongly reduced and number of equilibrium structures grows exponentially with the number $N$ of particles in the structure [24]. The list of spherical TSs with the lowest ever seen energy is constantly updated [23] by Bowick group of physicists.

Note that some of the TSs with N<400 demonstrate icosahedral symmetry [22] and up to the numerical optimization correspond to the Caspar and Klug (CK) structures of viral capsids [3]. Here, we adapt and develop the initial CK geometrical model to search for the lowest-energy TSs which have slightly broken icosahedral symmetry and numbers of particles prohibited in the CK capsid model. These TSs appear to be much more spread and our approach allows replacing about 10% of structures from the list [23] in the interval 600≤$N$≤1000. We analyze this interval since the related arrangements of particles are sufficiently complicated, but simultaneously these crystals are well investigated by other authors.

Large ($N$>400) known TSs are simply connected hexagonal lattices with 12 extended topolological defects [23] induced by the sphere curvature. Most of spherical viral capsids are also described in terms of hexagonal lattices [3, 25]. Quasiequivalence [3] of proteins in capsids makes their global symmetry the icosahedral one. However, icosahedral arrangements of particles are possible only at the particular $N$ values. Moreover, even for these $N$ values there is not any physical reason (like entropy contribution into free energy) which favors the high symmetry of sufficiently large TSs. Therefore, slight distortions of icosahedral lattices can simultaneously increase admitted $N$ values and make the distorted structures energetically favorable. Following this idea we constructed and optimized several hundred of trial structures from the interval 600≤$N$≤1000 during a few hours of calculation with laptop. As a result of this work for 40 values of $N$ we found new spherical crystals with energies lower than those of earlier seen energetically best structures. All the TSs found belong to the same new type of the ground states, where the minimal-size topological defects are the identical flattened pentagons and their vacant geometric centers are located at the vertices of regular or slightly deformed spherical icosahedron.

The proposed below geometric method to construct spherical hexagonal lattices with slightly broken icosahedral symmetry doesn't depend on the explicit form of the particles coupling. However, to compare our results with the previous ones we put $\alpha=1$ and choose the energy $F$ of the particles interaction in the form:

$$F = \sum_{j>i}^{N} \frac{1}{r_{ij}}, \qquad (1)$$

Let us recall that sufficiently large TSs with ($N$>400) demonstrate the local hexagonal order and contain 12 extended topological defects [6, 26] (ETDs), located near the vertices of regular or slightly irregular icosahedron covered by a simply connected hexagonal lattice. The ETDs are induced by curvature and the hexagonal order in their areas is strongly distorted. Each ETD carries a total topological charge +1. After a triangulation the ETDs are presented as linear scars [4] consisting of closely located particles with different surroundings. Particles having 5 or 7 nearest neighbors alternate in the scars, therefore they are usually treated as sequences of elementary 5-fold and 7-fold disclinations. Note that like the TSs the spherical colloidal crystals also possess 12 scars located similarly [26].

Comparison between spherical structures corresponding to more or less deep energy minima of Eq. (1) shows [20] that when the spherical hexagonal order tends to its ground state, it becomes more perfect, the areas occupied by the extended defects become smaller, and the ETD structures becomes simpler. Nevertheless, the limit of this simplification representing a point disclination is not achieved and the lowest energy spherical structures cannot be reduced to the CK icosahedral lattices having 12 point topological defects.

According to the CK theory [3], the structures of a large number of spherical viral capsids is interpreted in terms of the icosahedron net decorated by a periodic hexagonal lattice with nodes containing capsomers constructed from proteins. The edges of the icosahedron assembled from the net are hexagonal translations, which are symmetrically equivalent to each other. Only two indexes ($h,k$) are sufficient to determine the icosahedron edges and to distinguish between the different capsid structures. The geometry of CK model provides the maximal equivalence of proteins in the viral capsid and imposes a restriction $10T+2$ on the total number of capsomeres in it. The triangulation number $T$ is the square of the icosahedron edge length: $T=h^2+k^2+hk$. The capsomers can contain five or six proteins. Each capsid includes 12 pentamers located in the icosahedron vertices. The pentamers in the model are the isolated topological defects, since they have not 6, but 5 nearest neighbors. The other capsomers are hexamers and their number is $10(T-1)$.

Note, that in spite of huge advances in the interpretation of the viral capsids structures [25], the use of original CK geometric model to generate trial TSs is very limited. The icosahedral arrangements of particles are possible at the particular $N$ values only. Moreover, the largest TS with the lowest ever seen energy [23], which yet correspond to the CK model, contains $N=392$ particles and is indexed as (2,5). However, as it is shown below, some simple and clear modifications of the CK model make it suitable for larger structures and simultaneously increase essentially the list of admitted $N$ values.

To introduce our approach, let us note that in larger spherical structures the ETDs decrease energy (1) because they reduce increased density of particles appearing on the sphere near the projections of the icosahedron vertices. To obtain new 40 most favorable energetically structures in the region $600 \leq N \leq 1000$ it is sufficient to introduce identical simple topological defects by exclusion of particles in the vertices of regular or *specifically irregular* icosahedra. To obtain the lowest-energy structures with $N>1000$ particles the more complicated ETDs arising due to exclusion of a larger number of particles near the icosahedrons vertices may be used. In a similar way several energetically favorable large structures with icosahedral symmetry were constructed [6]. Below we consider the case of regular icosahedra and then we pass to the problem how to construct energetically favorable structures based on slightly irregular icosahedral arrangements of particles.

Exclusion of 12 nodes in vertices of CK icosahedrons leads to a series of trial structures indexed like capsids by integer ($h,k$) and containing $N=10(T-1)$ particles. We started our computations for fifteen trial structures corresponding to the interval $600 \leq N \leq 1000$, which contains only fourteen different values of $N$ from this series, since $9^2+9\cdot1+1=6^2+6\cdot5+5^2$. Before the gradient decent we changed randomly and very slightly the initial coordinates of particles. It induces bifurcations between the energy (1) minima. The arising low-symmetry structures (see examples in Fig.1) have extremely close but different energies. The deepest energy minima are more frequent and about 10 variations of the initial conditions are sufficient to see the deepest minimum again. After the numerical optimization performed for all the trial structures we found 8 unknown earlier crystals with the lowest ever seen energies (see Table 1 and Supplementary materials) and one energetically best structure (containing 600 particles) already known [23]. Only one of the crystals found (originating from the trial structure with $I_h$ symmetry and N=740 particles) is invariant with respect to inversion. All the other ones are completely asymmetric.

The fact that the optimization breaks the icosahedral symmetry of the above TSs is not surprising since energy (1) is purely potential. In contrast with a conventional free energy it does not contain an entropy term $-TS$, which usually favors high symmetry crystals in the high-temperature region. When the temperature *T* decreases the most of crystals undergo phase transitions breaking their symmetry [27].

*Table 1. Parameters of the lowest-energy spherical structures obtained from the net of regular icosahedra with excluded vertices. The first column gives the number of particles in the structure while the second one represents icosahedron indices.*

| $N$ | $(h,k)$ | Energy (1) obtained here | Energy (1) from Ref. 23 |
|-----|---------|--------------------------|--------------------------|
| 620 | (3,6)   | 183673.0995320           | 183673.1006540           |
| 660 | (2,7)   | 208434.7811871           | 208434.8462991           |
| 740 | (5,5)   | 262680.3895493           | 262680.3910595           |
| 750 | (4,6)   | 269904.2262369           | 269904.2627341           |
| 780 | (3,7)   | 292166.7658372           | 292166.8010876           |
| 900 | (5,6)   | 390084.7556863           | 390084.7935369           |
| 920 | (4,7)   | 407784.8388323           | 407784.8649502           |
| 960 | (3,8)   | 444368.8193481           | 444368.8463700           |

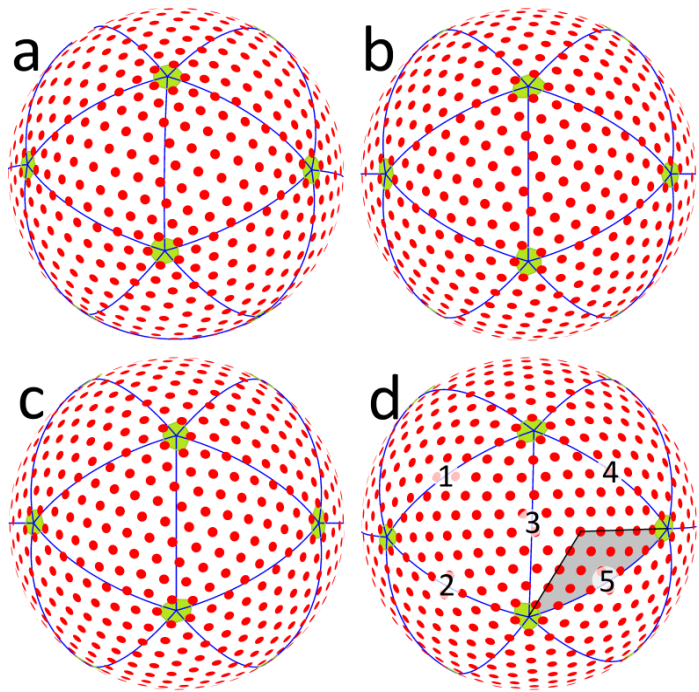

*Figure 1. Examples of structures arising due to exclusion of nodes in the icosahedrons vertices and containing N=10(T-1) particles. (a): Ideal spherical CK (7,3) icosahedron with 12 vacant positions in its vertices. (b-c): In the optimized structures regular pentagons surrounding the vertices become flatten and the icosahedral symmetry is broken. The structures differ from each other by the flattened pentagons orientations and are extremely close in energy. The spherical icosahedra (shown in blue) preserve the regular shape since the optimization does not shift the ETDs centers. (d): the lowest-energy structure with N = 800 obtained in Ref. 23 is based on an irregular icosahedron with different edges. Triangulation indices of the edges enumerated by numbers from 1 to 5 are (5,6), (5,5), (5,6), (4,6), (5,5), respectively. These indices are determined by the side lengths of triangulation triangles. For the 5$^{th}$ edge, this triangle is highlighted with gray.*

Thus, we have analyzed the energies of all 14 spherical crystals in the region $600 \leq N \leq 1000$ obtained from regular CK structures by removing 12 vertices and subsequent optimization. 8 of them turn out to have the lowest ever seen energy, one structure has the same energy as that reported in [23]. Last five structures obtained by this method have the higher energies than arrangements found earlier [23]. Four of these arrangements possess the ETDs in the form of scars. The last structure with $N=800$ is presented in Fig. 1(d) and looks like those presented in panels 1(b-c). Note, that $800=10(81-1)$. Nevertheless, the icosahedron corresponding to the structure is irregular and the triangulation number $T=81$ cannot be assigned to it. Spherical crystal presented in Fig. 1(d) and similar crystals with other $N$ values can be energetically favorable since the coupling between the ETDs with the flatten shape differs from the interaction between the point defects. Even in the case when the number $N$ of particles is compatible with the location of ETDs in the regular icosahedron vertices, the structure corresponding to a slightly distorted icosahedron may have smaller energy.

To proceed further and to study spherical structures, based on the irregular icosahedra, let us discuss how to obtain a simplest distortion of the regular icosahedron net. The problem is complicated by the condition that the distorted icosahedron should be smoothly covered by the hexagonal order. Repulsion of topological defects suggests that the icosahedron distortion should be small. Figure 2 shows an example of the (5,5) regular icosahedron net drawn on the hexagonal lattice, and demonstrates how one can deform the net in the simplest way.

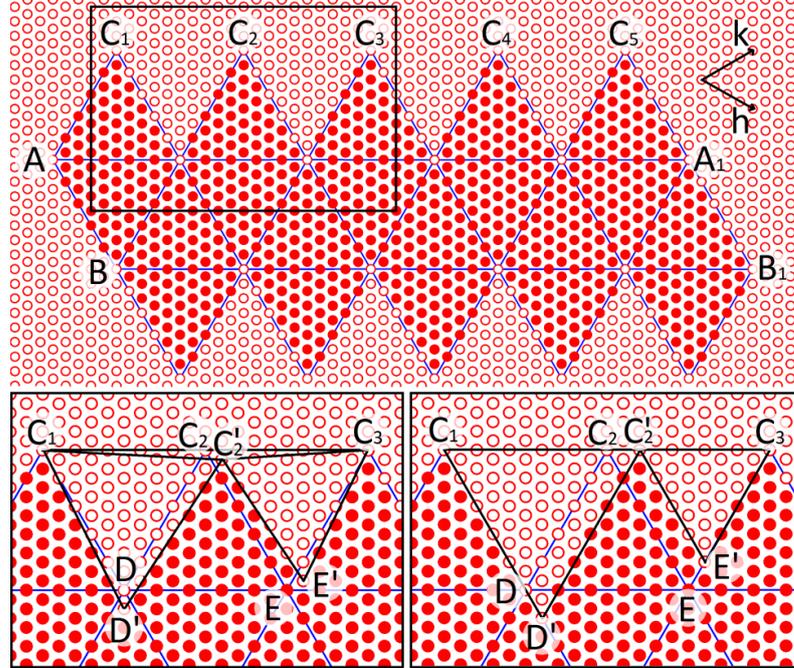

*Figure 2. The net of regular icosahedron (5,5) drawn on a hexagonal lattice. Two insets in the bottom of the figure show two cases of the simplest distortion of the region bounded by the rectangle. Point $C_2$ is shifted by arbitrary lattice translation $(h',k')$, its new position is $C_2'$. To ensure the smooth conjugation of hexagonal order at adjacent faces the vertices D and E must also be shifted by the translations $(h'+k', -h')$ and $(-k', h'+k')$, their new positions are D' and E', respectively. In the left inset: $(h',k') = (1,0)$ and in the right one: $(h',k') = (1,1)$.*

Geometrical reasons impose strong restriction upon the icosahedron net drawn on a hexagonal lattice. Before and after its distortion all the net vertices should coincide with the lattice nodes. Smooth conjugation of hexagonal order at the adjacent faces of icosahedron implies additional conditions on the shift of the net vertices. In particular, the translations $(AB)$ and $(A_1B_1)$ should remain equal and parallel to each other. In addition, before and after the distortion all eliminated from the net sectors (like $C_1D'C_2'$) should be of $60^0$. Any shift of a single vertex of the net is impossible. The net distorted in this way cannot be assembled into any icosahedron. However, one can freely shift one vertex of the net if the neighboring vertices are shifted accordingly. The smallest number of compensatory shifts (only two) is required to move the net vertex, which after icosahedron assemblage coincides with four other one of the same type. These vertices are noted as $C_i$ in Fig. 2, where the selected vertex $C_2$ is shifted by an arbitrary translation $(h',k')$. The neighboring vertices D and E should be shifted by the compensating translations $(h'+k',-h')$ and $(-k', h'+k')$, respectively. Calculation of areas of three triangles $(C_1D'C_2')$, $(C_2'E'C_3)$ and $(C_1C_2'C_3)$ determines the change of the particles number $\Delta N$ in the distorted icosahedron:

$$\Delta N = -T' + (hk' - h'k), \qquad (2)$$

where $T' = (h'^2 + h'k' + k'^2)$ is the second triangulation number, and the last term with the alternating sign in (2) is associated with the area of the triangle $(C_1C_2'C_3)$.

It is interesting to discuss some geometrical properties of the irregular icosahedrons obtained. The above net reconstruction changes 8 from 20 of icosahedron faces. These distorted faces contain the vertices of the edge (D'E'). The other 12 faces preserve the initial regular shape. The edge (D'E') is shifted and slightly rotated with respect to its initial location (DE). Therefore all symmetry elements of the initial regular icosahedron disappear. However, due to the particular form of the icosahedron distortion a new two-fold axis appears. It is directed from the middle of (D'E') edge to the middle of the opposite undistorted edge. This is the single symmetry element of the irregular icosahedrons under consideration. Thus, the exact crystallographic symmetry of these icosahedrons is always $C_2$ independently on values of $(h,k)$ and $(h',k')$ vectors. However, an approximate (slightly broken) icosahedral symmetry is obviously manifested in all the structures considered in this Letter.

The introduced above simplest distortion of the icosahedron net is universal and different slightly irregular icosahedra may be constructed with its help. For example, combining a couple of simplest distortions it is easy to obtain the structures with $\Delta N = -2T'$ or consider another class of trial spherical structures, preserving inversion. In this article we do not consider such multiple distortions and focus on the simplest case.

Note, that for each $T'$ value there are six inequivalent simplest distortions of the regular net. They are obtained by consequent $60^0$ rotations of the translation $(h',k')$. It is interesting to note that for a regular net (4,6) all six variants of the simplest distortions with $(T'=1)$ lead to the earlier unknown energetically best structures (see Table. 2 and Ref. 23).

*Table 2. Parameters of the previously unknown energetically best structures obtained on the basis of simplest distortions of the (4,6) regular icosahedron with excluded vertices. The first column gives the number of particles in the structure, the second one contains the shift vector $(h',k')$.*

| N | $(h',k')$ | Energy (1) for our new structures | Energy (1) from Ref. 23 |
|---|---|---|---|
| 739 | (1,-1) | 261964.004594769 | 261964.213759900 |
| 743 | (1,0) | 264837.700971304 | 264838.220062800 |
| 745 | (0,-1) | 266280.731191900 | 266280.417488466 |
| 753 | (0,1) | 272090.808556541 | 272091.610155500 |
| 755 | (-1,0) | 273553.413036802 | 273553.821920500 |
| 759 | (-1,1) | 276490.454788544 | 276490.484843900 |

Considering all possible simplest distortions of regular icosahedron structures with eliminated vertices, we found 31 new energetically best structures with $T'=1$ and one energetically best arrangement with $T'=3$. The energetically favorable structures are obtained, only if $T'\leq3$. The variant $T'=2$ is prohibited due to geometric reasons.

Fig. 3 shows the construction and optimization of two new lowest energy spherical structures. Their nets are presented in Fig.2. Note that in contrast to the structures based on the regular icosahedron net, optimization of those based on the net of the irregular icosahedron is mostly unambiguous. Small changes in initial positions of particles in 30 trial structures from the 32 ones do not lead to switching between minima of energy (1), since the $C_2$ symmetry does not change during their optimization. The optimization of two trial structures (N=711 and N=894) breaks the $C_2$ symmetry. However, only in the latter structure small changes in the initial positions lead to switching between the minima with very close energies.

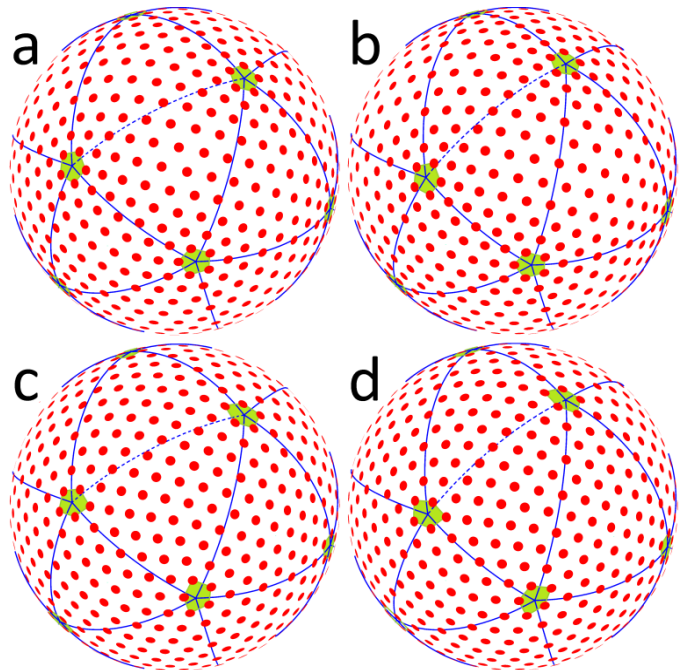

*Figure 3. Examples of trial (panels (a,b)) and optimized (panels (c,d)) spherical crystals with N=734 (left column) and N=737 (right column). The nets of these structures are shown in figure 2. The crystals in panels (c-d) present two from 32 new energetically best structures based on the irregular icosahedra with excluded 12 vertices. Their energies are lower than the lowest earlier seen energy. Two-fold axes (see the text) intersect the centers of edges shown by dashed lines.*

The energies for all 40 new energetically best (to date) structures found in this article are presented in Supplementary materials.

In conclusion, Thomson problem (search for the lowest-energy structures formed by repelling particles retained on a spherical surface) arises on different levels of the matter self-organization. Here we advance the problem proposing the new physically justified type of trial spherical structures for a subsequent numerical optimization, which is simple since our trial structures are close to the lowest energy ones. The found crystals with the lowest ever seen energy possess the curvature induced topological defects being flatten pentagons with the centers located in the vertices of perfect or slightly distorted spherical icosahedrons covered by perfect hexagonal order. Similar scars-free type of compatibility between hexagonal order and spherical geometry may be very common among the TSs. Our analysis shows that in the considered interval of *N* values 30 known TSs [23] demonstrate it in addition to 40 structures obtained here. The list of analogous TSs may be expanded by considering trial structures, where the simplest distortions of the global icosahedral order are combined. Following this way one can obtain the energetically best structures with arbitrary numbers of particles.

Our results may be interesting for physicists working on theoretical and experimental problems of self-assembly in various types of spherical nano- and micro-structures. The most of these 2D shells is characterized by the local hexagonal order of different types and global (perfect or slightly distorted) icosahedral arrangement. In the frame of the Thomson problem this article demonstrates how the simplest energetically favorable distortion of the global icosahedral arrangements occurs. However, let us stress, that this distortion was found from the geometrical consideration and, consequently, it is independent on the type of specific interaction between the particles forming the shell. Moreover, since we consider the distortions of the *lattices* covering the sphere, our results are also independent on the particular type of the local order implemented in the spherical structure. Therefore the structures with the above considered simplest distortions can be discovered in course of further experimental investigations of misassembled viral capsids or fullerenes.

**Acknowledgements**

Authors acknowledge financial support of the RFBR grant 13-02-12085 ofi_m.

**Supplementary material**

***Supplementary table***. *Parameters of the lowest-energy spherical structures obtained from the nets of regular and slightly irregular icosahedra with excluded vertices. The first column gives the number of particles in the structure, the second and the third one contain (h, k) and (h', k') vectors, respectively. The fourth column gives the difference ΔN (see Eq. 2) between the number of particles in regular and slightly irregular icosahedral structures. The point symmetry groups of the crystals found are listed in the fifth column: $C_1$ denotes the trivial symmetry, $C_2$ is the group generated by two-fold axis, and I is the group generated by inversion. The last two columns contain the energy of our structures and that of the lowest seen earlier structures [23] with the same numbers of particles.*

| N | (h,k) | (h',k') | ΔN | Point Group | Energy (1) for our new structures | Energy (1) for the best TSs from Ref. 23 |
|---|---|---|---|---|---|---|
| 604 | (4,5) | (-1,0) | 4 | $C_2$ | 174209.2950048 | 174209.3433670 |
| 616 | (3,6) | (0,-1) | -4 | $C_2$ | 181283.7994340 | 181283.8637811 |
| 620 | (3,6) | (0,0) | 0 | $C_1$ | 183673.0995320 | 183673.1006540 |
| 622 | (3,6) | (0,1) | 2 | $C_2$ | 184873.9213455 | 184873.9706900 |
| 657 | (2,7) | (0,-1) | -3 | $C_2$ | 206523.2960746 | 206523.4569710 |
| 660 | (2,7) | (0,0) | 0 | $C_1$ | 208434.7811871 | 208434.8462991 |
| 661 | (2,7) | (0,1) | 1 | $C_2$ | 209074.0693270 | 209074.2722963 |
| 711 | (1,8) | (1,0) | -9 | $C_1$ | 242289.3483748 | 242289.3505490 |
| 734 | (5,5) | (1,0) | -6 | $C_2$ | 258393.8379678 | 258394.0288507 |
| 737 | (5,5) | (1,1) | -3 | $C_2$ | 260533.5191931 | 260533.5975723 |
| 739 | (4,6) | (1,-1) | -11 | $C_2$ | 261964.0045947 | 261964.2137599 |
| 740 | (5,5) | (0,0) | 0 | I | 262680.3895493 | 262680.3910595 |
| 743 | (4,6) | (1,0) | -7 | $C_2$ | 264837.7009713 | 264838.2200628 |
| 744 | (5,5) | (-1,0) | 4 | $C_2$ | 265558.4086906 | 265558.7257592 |
| 745 | (4,6) | (0,-1) | -5 | $C_2$ | 266280.4174884 | 266280.7311919 |
| 750 | (4,6) | (0,0) | 0 | $C_1$ | 269904.2262369 | 269904.2627341 |
| 753 | (4,6) | (0,1) | 3 | $C_2$ | 272090.8085565 | 272091.6101555 |
| 755 | (4,6) | (-1,0) | 5 | $C_2$ | 273553.4130368 | 273553.8219205 |
| 759 | (4,6) | (-1,1) | 9 | $C_2$ | 276490.4547885 | 276490.4848439 |
| 772 | (3,7) | (1,0) | -8 | $C_2$ | 286143.8799870 | 286143.9594358 |
| 776 | (3,7) | (0,-1) | -4 | $C_2$ | 289147.5579677 | 289147.7151616 |
| 780 | (3,7) | (0,0) | 0 | $C_1$ | 292166.7658372 | 292166.8010876 |
| 782 | (3,7) | (0,1) | 2 | $C_2$ | 293682.6024071 | 293682.7579402 |
| 786 | (3,7) | (-1,0) | 6 | $C_2$ | 296725.9510131 | 296726.2156791 |
| 821 | (2,8) | (1,0) | -9 | $C_2$ | 324026.6761071 | 324026.8643390 |
| 827 | (2,8) | (0,-1) | -3 | $C_2$ | 328827.8951167 | 328828.0997547 |

| 831 | (2,8) | (0,1) | 1 | $C_2$ | 332048.3014829 | 332048.3684809 |
| 837 | (2,8) | (-1,0) | 7 | $C_2$ | 336908.8648947 | 336908.9040403 |
| 893 | (5,6) | (1,0) | -7 | $C_2$ | 383983.4934817 | 383983.8873440 |
| 894 | (5,6) | (0,-1) | -6 | $C_1$ | 384852.1843044 | 384852.7091227 |
| 900 | (5,6) | (0,0) | 0 | $C_1$ | 390084.7556863 | 390084.7935369 |
| 904 | (5,6) | (0,1) | 4 | $C_2$ | 393593.5478252 | 393593.6512960 |
| 905 | (5,6) | (-1,0) | 5 | $C_2$ | 394473.2029360 | 394473.6617797 |
| 910 | (5,6) | (-1,1) | 10 | $C_2$ | 398886.1796926 | 398886.1869390 |
| 920 | (4,7) | (0,0) | 0 | $C_1$ | 407784.8388323 | 407784.8649502 |
| 923 | (4,7) | (0,1) | 3 | $C_2$ | 410474.1375083 | 410474.8997608 |
| 926 | (4,7) | (-1,0) | 6 | $C_2$ | 413172.2118282 | 413172.5104270 |
| 930 | (4,7) | (-1,1) | 10 | $C_2$ | 416783.4581803 | 416783.6980906 |
| 960 | (3,8) | (0,0) | 0 | $C_1$ | 444368.8193481 | 444368.8463700 |
| 967 | (3,8) | (-1,0) | 7 | $C_2$ | 450933.6496672 | 450933.7609188 |